\documentclass[aps,prb,preprint,groupedaddress]{revtex4}

\usepackage{graphicx}
\usepackage{bm}

\begin{document}

\title{The effect of in-plane magnetic field on the spin Hall effect
in Rashba-Dresselhaus system}

\author{Ming-Che Chang}
\affiliation{Department of Physics, National Taiwan Normal
University, Taipei, Taiwan}

\date{\today}
\begin{abstract}
In a two-dimensional electron gas with Rashba and Dresselhaus
spin-orbit couplings, there are two spin-split energy surfaces
connected with a degenerate point. Both the energy surfaces and
the topology of the Fermi surfaces can be varied by an in-plane
magnetic field. We find that, if the chemical potential falls
between the bottom of the upper band and the degenerate point,
then simply by changing the direction of the magnetic field, the
magnitude of the spin Hall conductivity can be varied by about 100
percent. Once the chemical potential is above the degenerate
point, the spin Hall conductivity becomes the constant $e/8\pi$,
independent of the magnitude and direction of the magnetic field.
In addition, we find that the in-plane magnetic field exerts no
influence on the charge Hall conductivity.
\end{abstract}

\maketitle

In spintronics devices, injection of spins efficiently from
ferromagnetic leads has remained a very challenging problem.
Therefore, alternative approaches to generate spin polarization
current are being intensively pursued. For example, one can use an
external electric field to manipulate the spin transport via
spin-orbit coupling. For bulk semiconductors, Murakami {\it et al}
showed that one can employ the spin-orbit coupling in hole bands
to generate a spin Hall current.\cite{murakami03} According to
them, the spin Hall current is dissipationless and related to the
Berry curvature, which usually vanishes in materials with both
inversion and time-reversal symmetry, such as Si, but could be
nonzero because of the spin-orbit coupling.

In two-dimensional electron gas (2DEG), spin-orbit coupling could
arise because of structure inversion asymmetry (the Rashba
mechanism),\cite{bychkov84} bulk inversion asymmetry (the
Dresselhaus mechanism),\cite{dresselhaus55} or other
mechanisms.\cite{zutic04} The existence of Rashba coupling in
heterojunctions has led to many creative proposals for its
applications. This includes, for example, the current modulator
proposed by Datta and Das,\cite{datta90} or the spin filter based
on electron focusing\cite{focusing}. The effect of Rashba coupling
in quantum wire,\cite{wire} quantum ring,\cite{ring} or quantum
dot\cite{dot} has also been investigated.\cite{negative} Recently,
Sinova {\it et al} showed that in a clean and infinite Rashba
system, one could generate a spin Hall current by an electric
field.\cite{sinova04} In a clean, free-electron-like system,
inclusion of both Rashba and Dresselhaus mechanisms still yield
the same (up to a sign change) spin Hall conductivity (SHC)
$e/8\pi$, independent of spin-orbit coupling strength and carrier
density if both bands are populated.\cite{sinitsyn04,shen04}

The robustness of the value $e/8\pi$ against factors such as
disorder, finite-size effect, and electron-electron interaction is
being actively studied. It is generally believed that strong
disorder would destroy the spin Hall effect in 2DEG. According the
some analysis, the $e/8\pi$ value could be preserved in weak
disorder.\cite{burkov03} However, several perturbative
calculations conclude that the spin Hall effect would disappear as
long as disorder exists.\cite{disorder} Numerical calculations
thus far tend to show that the SHC is indeed robust against weak
disorder, but reduces to zero at strong disorder.\cite{numerical}
This issue remains to be clarified. Besides disorder, it is found
that the interactions between electrons could renormalize the SHC
to some extent.\cite{dimitrova04,rashba}

For the 2DEG with both Rashba and Dresselhaus spin-orbit
couplings, there are two energy surfaces connected with a
degenerate point at momentum ${\vec k}_0={\vec 0}$. The SHC is a
constant when both bands are populated.\cite{sinitsyn04,shen04} If
the chemical potential $\mu$ is below the minimum energy of the
upper band (denoted by $E_{+,{\rm min}}$, which equals the
degenerate energy $E_*$ in this case), the SHC would depend on
electron density.\cite{sinova04} In the presence of an in-plane
magnetic field, the energy surfaces are distorted such that ${\vec
k}_0\neq {\vec 0}$ and $E_{+,{\rm min}}$ could be below $E_*$. It
would be interesting to investigate the effect of distortion on
the SHC. Our major finding is that, when $\mu$ is lying between
$E_{+,{\rm min}}$ and $E_*$, one can vary the magnitude of the SHC
by as much as 100 percent simply by changing the direction of the
magnetic field. When $\mu$ is above $E_*$ (rather than $E_{+,{\rm
min}}$), the SHC retains the same constant $e/8\pi$, independent
of the direction and magnitude of the magnetic field. Even though
the in-plane magnetic field has significant effect on the SHC, it
exerts no influence on the charge Hall conductivity, which remains
zero as the topology of (one-dimensional) Fermi surfaces are
changed due to the rising chemical potential.

We consider the following Hamiltonian,
\begin{equation}
H=\frac{p^2}{2m^*}+\frac{\alpha}{\hbar}(\sigma_x p_y-\sigma_y
p_x)+\frac{\gamma}{\hbar}(\sigma_x p_x-\sigma_y
p_y)+\beta_x\sigma_x +\beta_y\sigma_y,
\end{equation}
in which $\alpha$ and $\gamma$ represent the Rashba coupling and
the Dresselhaus coupling, $\vec{\beta}=(g^*/2)\mu_B {\vec B }$,
where $g^*$ is the effective $g$-factor (assumed isotropic and
field-independent) and ${\vec B}$ is the in-plane magnetic field.
The finite thickness of the 2DEG layer has been neglected so the
magnetic field couples only to the electron spin. The
eigen-energies of the Hamiltonian are
\begin{equation}
E_\lambda({\vec k})=E_0({\vec k})+\lambda\sqrt{(\gamma k_x+\alpha
k_y+\beta_x)^2+(\alpha k_x+\gamma k_y-\beta_y)^2},
\label{energies}
\end{equation}
where $E_0({\vec k})=\hbar^2 k^2/2m^*$ and $\lambda=\pm$, with the
corresponding eigen-states,
\begin{equation}
|{\vec k},+\rangle=\frac{1}{\sqrt{2}}\left(\begin{array}{c}
1\\-ie^{i\theta}\end{array}\right); |{\vec
k},-\rangle=\frac{1}{\sqrt{2}}\left(\begin{array}{c}
-ie^{-i\theta}\\1\end{array}\right), \label{states}
\end{equation}
where $\tan\theta=(\gamma k_x+\alpha k_y+\beta_x)/(\alpha
k_x+\gamma k_y-\beta_y)$. There are several legitimate but
different choices of the eigenstates. Usually, the exponential
factors are placed at the same lower (or upper) position for the
two spinors. The choice in Eq.~(\ref{states}) ensures that the
eigenstates are free of phase ambiguity at the degenerate point
${\vec k}_0$ in the presence of an infinitesimal and perpendicular
magnetic field.\cite{sheng96} This choice is crucial in obtaining
the correct Hall conductance, as will be explained in more details
later.

It can be seen from Eq.~(\ref{energies}) that the two energy
surfaces are degenerate at the point ${\vec
k}_0=((\gamma\beta_x+\alpha\beta_y)/(\alpha^2-\gamma^2),
-(\alpha\beta_x+\gamma\beta_y)/(\alpha^2-\gamma^2))$. When
$\alpha=\gamma$, there is no degeneracy in general. But if
$\beta_x$ also equals $\beta_y$, then there is a line degeneracy
along $k_x-k_y=\beta_x/\alpha$, similar to the case for ${\vec
\beta}={\vec 0}$.\cite{sinitsyn04,shen04} The energy contours for
energies located below, at, and above the degenerate point are
shown in Fig.~1.\cite{below} The degenerate point is moving from
outside of the contours to inside as the energy $E$ is increasing.
We find that the density of states $g(E)$ undergoes a finite jump
when $E$ is crossing a band bottom, and it can be proved that
$g(E)=m/(\pi\hbar^2)$, same as the free electron value, as long as
$E$ is above $E_*=E({\vec k}_0)$.

\begin{figure}
\includegraphics[width=4.5in]{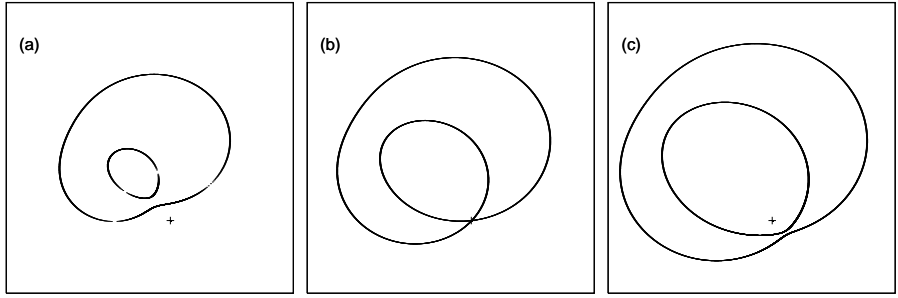}
\caption{Schematic diagrams of the energy contours for energies
located (a) below, (b) at, and (c), above the degenerate energy
$E_*$. The degenerate point ${\vec k}_0$ is marked with a cross.
There is no rotational symmetry for the contours when $\alpha$,
$\gamma$, and ${\vec \beta}$ are all nonzero.}
\end{figure}

The spin Hall conductivity is given by the Kubo
formula,\cite{sinova04}
\begin{equation}
\sigma_{\mu\nu}^\eta=\frac{1}{i\hbar}\sum_{\vec
k}\sum_{\lambda,\lambda'(\lambda\neq\lambda')} \frac{f_{{\vec
k},\lambda}-f_{{\vec
k},{\lambda'}}}{\omega_{\lambda\lambda'}^2({\vec k})} \langle{\vec
k},\lambda|j_\mu^\eta|{\vec k},\lambda'\rangle\langle{\vec
l},\lambda'|j_\nu|{\vec k},\lambda\rangle, \label{skubo}
\end{equation}
where $j_\mu^\eta=(\hbar/4)(v_\mu\sigma_\eta+\sigma_\eta v_\mu)$
is the generally accepted definition of the spin current,
$j_\nu=-e v_\nu$ is the electric current, $v_\mu=\partial{\tilde
H}({\vec k})/\partial(\hbar k_\mu)$, ${\tilde H}({\vec k})\equiv
e^{-i{\vec k}\cdot{\vec r}}He^{i{\vec k}\cdot{\vec r}}$, and
$\hbar\omega_{\lambda\lambda'}({\vec k})\equiv E_\lambda({\vec
k})-E_{\lambda'}({\vec k})$. It can be shown that both
$\sigma_{xy}^x$ and $\sigma_{xy}^y$ are zero in the absence of
disorder, and
\begin{equation}
\sigma_{xy}^z=-\frac{2e}{m\hbar}\sum_{\vec k}(f_{{\vec
k},-}-f_{{\vec
k},+})\frac{k_x^2(\alpha^2-\gamma^2)-k_x(\alpha\beta_y+\gamma\beta_x)}
{\omega_{+-}^3},\label{main}
\end{equation}
where $\hbar\omega_{+-}=2[(\alpha k_x+\gamma
k_y-\beta_y)^2+(\gamma k_x+\alpha k_y+\beta_x)^2]^{1/2}$. A
typical result for the SHC as a function of energy is plotted in
Fig.~2. For the special case of $\gamma=0$, the degenerate point
${\vec k}_0={\vec \beta}\times\hat{z}/\alpha$, and
$E_*=(\hbar^2/2m^*)(\beta/\alpha)^2$. For $\beta>\alpha^2/2$,
which is the case for Fig.~2, the bottoms of upper and lower bands
are at $E_{\pm,{\rm min}}=-(m^*\alpha/\hbar^2)^2\pm\beta$.
Numerical values of these energies are given in the figure
caption. It can be seen that the SHC becomes nonzero when $E$ is
above the bottom of the lower band. It remains monotonic when $E$
is between the bottoms of lower and upper bands, and may rise over
the value $e/8\pi$. For most angels of the magnetic field, the
abrupt turn of $\sigma_{xy}^z$ at $E=E_{+,{\rm min}}$ forms a
cusp. Once the energy reaches $E_*$, the SHC becomes the constant
$e/8\pi$, independent of the direction of ${\vec \beta}$.

Notice that, according to our choice, the transverse spin current
is along the $x$-direction and the longitudinal electric field is
along the $y$-direction. When the energy $E$ equals $E_{+,{\rm
min}}$, the magnitude of $\sigma_{xy}^z$ can be changed by roughly
100 percent when ${\vec B}$ rotates from the transverse direction
(minimum) to the direction of the electric field (maximum). As the
angle increases further, the SHC again reaches minimum along the
$-x$ direction and rises to maximum along the $-y$ direction. That
is, when $\gamma=0$, there is a two-fold symmetry in the plot of
$\sigma_{xy}^z({\vec B})$.

\begin{figure}
\includegraphics[width=3.5in]{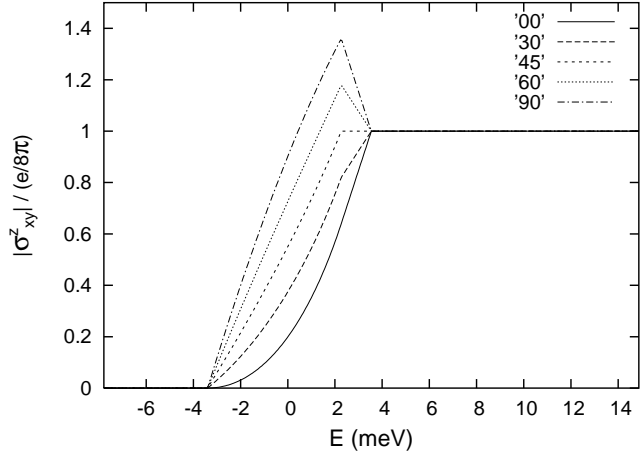}
\caption{ The spin Hall conductivity as a function of energy for
different directions of the magnetic field. The angles between
${\vec B}$ and $x$-axis are 0, 30, 45, 60, and 90 degrees. The
parameters are $m^*=0.024 m_e$, $\alpha=6\times
10^{-9}$eV$\cdot$cm, $\gamma=0$, $B=6.52$ T ($g^*=15$). The
bottoms of lower/upper bands and the degenerate energy are at
$E_{-,{\rm min}}=-3.4$ meV, $E_{+,{\rm min}}$=2.27 meV and
$E_*$=3.54 meV. }
\end{figure}

In our model (Eq.~(1)), in the most general cases when both
$\alpha$ and $\gamma$ are nonzero, the SHC still remains the
constant $e/8\pi$ when $E$ is above $E_*$. We sketch the proof
below to show that this is indeed true for all ranges of the
parameters $\alpha,\gamma$ and ${\vec \beta}$.\cite{exclude} While
evaluating the integral in Eq.~(\ref{main}), it is more convenient
to shift the origin to ${\vec k}_0$. Then the ${\vec
\beta}$-dependence of $\omega_{+-}$ in the denominator can be
eliminated,
\begin{equation}
\hbar\omega_{+-}=2k\left(\alpha^2+\gamma^2
+2\alpha\gamma\sin2\phi\right)^{1/2}\equiv2kg(\phi),
\end{equation}
where the magnitude $k$ and polar angle $\phi$ are relative to the
new origin ${\vec k}_0$. The $k$-integration now can be carried
out analytically,
\begin{eqnarray}
\sigma_{xy}^z
&&=-\frac{e\hbar^2}{4m}\frac{\alpha^2-\gamma^2}{(2\pi)^2}
\int_0^{2\pi}d\phi\frac{(k_-(\phi)-k_+(\phi))\cos^2\phi}{g^3(\phi)} \nonumber \\
&&-\frac{e\hbar^2}{4m}\frac{\alpha\beta_y+\gamma\beta_x}{(2\pi)^2}
\int_0^{2\pi}d\phi\frac{\ln(k_-(\phi)/k_+(\phi))\cos\phi}{g^3(\phi)},
\label{integral}
\end{eqnarray}
where $k_\pm$ are the lower and upper bounds of the
$k$-integration. As long as the energy is above $E_*$, the values
of $k_-$ and $k_+$ are unique (see Fig.~1(c)),
\begin{eqnarray}
k_\lambda(\phi) &&=-k_0\cos(\phi-\phi_0)-\frac{\lambda}{2}g(\phi)
\nonumber \\
&&+\sqrt{[k_0\cos(\phi-\phi_0)+\lambda
g(\phi)/2]^2+E-E_*},\label{k}
\end{eqnarray}
where $\phi_0$ is the angle between ${\vec k}_0$ and the $x$-axis.
From Eq.~(\ref{k}), one may not expect that the SHC would be a
constant in energy, since both $k_-$ and $k_+$ depend on energy
$E$ explicitly. Nevertheless, we can show that the integrals in
Eq.~(\ref{integral}) are independent of energy. Substitute the
difference,
\begin{eqnarray}
k_-(\phi)-k_+(\phi) &&=g(\phi) \nonumber \\
&&+\sqrt{[k_0\cos(\phi-\phi_0)-g(\phi)/2]^2+E-E_*} \nonumber \\
&&-\sqrt{[k_0\cos(\phi-\phi_0)+ g(\phi)/2]^2+E-E_*}, \label{kdiff}
\end{eqnarray}
to the first integral in Eq.~(\ref{integral}), we find that the
first term $g(\phi)$ contributes $e/8\pi$ to the SHC. The first
square root in Eq.~(\ref{kdiff}) is equal to the second one after
the shift $\phi\rightarrow\phi+\pi$. Since both $\cos^2\phi$ and
$g(\phi)$ in Eq.~(\ref{integral}) are invariant under
$\phi\rightarrow\phi+\pi$, these two terms in Eq.~(\ref{kdiff})
would cancel with each other after integration. Similarly, for the
second term in Eq.~(\ref{integral}), one can show that the ratio
$k_-(\phi)/k_+(\phi)$ is invariant under
$\phi\rightarrow\phi+\pi$. Therefore, the second integral in
Eq.~(\ref{integral}) vanishes since the $\cos\phi$ in the
numerator changes sign after a $\pi$-rotation.

The proof above is valid for $E>E_*$. At the degenerate energy
$E=E_*$, the denominator $\omega_{+-}$ is zero and the proof above
does not apply, but it still can be proved that the SHC converges
to the value $e/8\pi$ without showing any singular behavior at
$E=E_*$.

\begin{figure}
\includegraphics[width=3.5in]{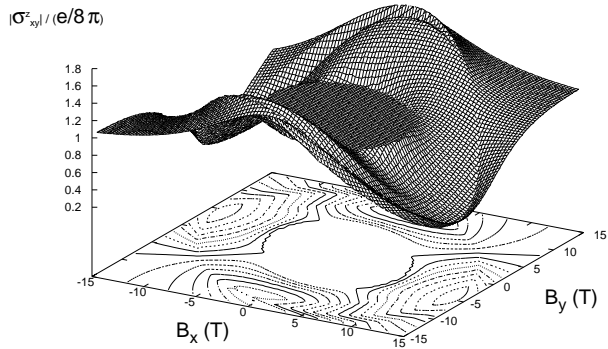}
\caption{Spin Hall conductivity plotted as a function of the
in-plane magnetic field ${\vec B}$. There is a plateau in the
middle when the magnetic field is weak. For stronger magnetic
field beyond the plateau region, a circle of cusp (ridge) is
clearly visible. Contours of the surface are projected onto the
plane below. Relevant parameters are specified in the text.}
\end{figure}

In Fig.~3, we show the SHC as a function of the in-plane magnetic
field ${\vec B}$ at a fixed electron density $n$. For smaller
magnetic field, the degenerate energy $E_*$ is lower than the
chemical potential. Therefore, there is a plateau with
$|\sigma_{xy}^z|=e/8\pi$ at the center of Fig.~3. The plateau has
an elliptical boundary defined by the equation $\mu=E_*$, or,
equivalently,
\begin{equation}
\beta_x^2+4\alpha\gamma\beta_x\beta_y+\beta_y^2=\zeta\frac{
(\alpha^2-\gamma^2)^2}{\alpha^2+\gamma^2},
\end{equation}
where $\zeta$ is a constant of the order of electron density. For
larger magnetic field, the degenerate point is driven out of the
Fermi sea and the SHC could be enhanced or reduced, depending on
the direction of ${\vec B}$ (similar to the behavior in Fig.~(2)).
The circle of cusp (ridge) surrounding the plateau is caused by
the crossing of $\mu$ and the bottom of the upper-band $E_{+,{\rm
min}}$ (see Fig.~2). Split cusps are visible in Fig.~3 in over
half of the ridge, which is due to the {\it re}-crossing of $\mu$
and $E_{+,{\rm min}}$. That is, when the magnetic field is
increasing at a fixed angle, because of the complex shift of the
bottoms of energy surfaces, the chemical potential $\mu(B)$ could
fall below $E_{+,{\rm min}}(B)$ after going through the first
ridge at $\mu(B_1)=E_{+,{\rm min}}(B_1)$, then rise to touch the
bottom of the upper band again at $\mu(B_2)=E_{+,{\rm min}}(B_2)$
($B_2>B_1$), where the second ridge (or cusp) is formed.

The Rashba and Dresselhaus couplings in Fig.~3 are $\alpha=6\times
10^{-9}$ eV$\cdot$cm and $\gamma=2\times 10^{-9}$
eV$\cdot$cm.\cite{rashba04} We choose the effective mass
$m^*=0.024 m_e$ and effective $g$-factor $g^*=15$ for electrons in
bulk InAs. The electron density $n$ is fixed at $5.7\times
10^{10}/{\rm cm}^2$ during the ${\vec B}$-scan. The ranges of
in-plane magnetic field are 15 Tesla in both directions. Based on
these realistic parameters, such angular variation of SHC might be
tested in future experiments. Different material parameters, such
as $m^*$ and $g^*$, could alter the range of appropriate $n$ and
$B$ by one or two orders of magnitude. In general, it is better to
choose materials with a large product of $m^*g^*$ to reduce the
magnetic field strength required.\cite{kittel}

At the end of the paper, we comment briefly on whether the
in-plane magnetic field could generate a charge Hall current in
the Rashba-Dresselhaus system. Replacing the spin current
$j_\mu^\eta$ in Eq.~(\ref{skubo}) by electric current $j_\mu$, the
Hall conductivity for the $\lambda$-band ($\lambda=\pm$) can be
written as\cite{takahashi96}
\begin{equation}
\sigma_{xy}^\lambda=\frac{e^2}{\hbar}\sum_{{\vec k}{\ }{\rm
filled}}\Omega_\lambda({\vec k}),\label{kubo}
\end{equation}
where the Berry curvature for the $({\vec k},\lambda)$-state is
\begin{equation}
\Omega_\lambda({\vec
k})=i\sum_{\lambda'\neq\lambda}\frac{\langle{\vec k},\lambda|v_x|
{\vec k},\lambda'\rangle\langle{\vec k},\lambda'|v_y|{\vec
k},\lambda\rangle- \langle{\vec k},\lambda|v_y|{\vec
k},\lambda'\rangle\langle{\vec k},\lambda'|v_x|{\vec
k},\lambda\rangle} {{\omega_{\lambda\lambda'}^2}({\vec
k})}.\label{curvature}
\end{equation}
It is easy to show that the Berry curvature is zero at every
${\vec k}$, except at the degenerate point where the formula above
does not apply. However, one can preform a line integral along an
arbitrary contour that encloses the degenerate point to obtain the
Berry phase,
\begin{equation}
\Gamma_\lambda=\oint d{\vec k}\cdot\langle{\vec
k},\lambda|i\frac{\partial}{\partial {\vec k}}|{\vec
k},\lambda\rangle=-{\rm
sign}(\alpha^2-\gamma^2)\lambda\pi,\label{loop}
\end{equation}
in which sign($\alpha^2-\gamma^2$)=$1, 0, -1$ when $\alpha^2$ is
larger than, equal to, or smaller than $\gamma^2$. This is
basically the same result as that of ${\vec \beta}=0$, and
reflects the fact that the Berry phase is topological in nature
and is not altered by the smooth distortion of the energy
surfaces. We note that the result in Eq.~(\ref{loop}) differs
slightly from that (for ${\vec \beta}=0$) in
Ref.~\onlinecite{shen04}, in which the signs of $\Gamma_+$ and
$\Gamma_-$ are the same (discuss below). From the non-zero Berry
phase, one can infer that the Berry curvature is singular at the
degenerate point, $\Omega_\lambda({\vec k})=-{\rm
sign}(\alpha^2-\gamma^2)\lambda\pi\delta({\vec k}-{\vec k}_0)$,
which is different from the usual monopole-field-like Berry
curvature. The result above is valid for the Rashba-Desselhaus
system in an arbitrary in-plane magnetic field.

Despite the non-zero Berry phase, the Hall conductance of the
system remains zero for whole range of the chemical potential. To
illustrate this point, let us consider the simpler ``pure" Rashba
system ($\gamma=0,{\vec \beta}=0$). If the chemical potential is
below the degenerate point, then only the lower band is populated
and the ${\vec k}$-integral in Eq.~(\ref{kubo}) is performed over
an annular region. The result of integration using
Eq.~(\ref{kubo}) is zero since the singularity of Berry curvature
is {\it not} located in the annulus. If the chemical potential is
above the degenerate point, then both bands are populated
(degenerate point included in {\it both} cases), and
$\sigma_{xy}=\sigma^+_{xy}+\sigma^-_{xy}=0$. Therefore, there
exists no charge Hall current no matter the chemical potential is
below or above the degenerate point. Different choices of bases in
Eq.~(\ref{states}) may lead to different results such as
$\sigma^\lambda_{xy}=e^2/(2h)$
($\lambda$-independent),\cite{shen04} which would give
$\sigma_{xy}=\sigma^+_{xy}+\sigma^-_{xy}=e^2/h$ when both bands
are populated. This would imply the counter-intuitive result that
the system exhibits charge Hall effect even in the small
spin-orbit coupling limit.\cite{gauge}

For the generic situation in which both $\gamma$ and ${\vec
\beta}$ are nonzero, the topology of the one dimensional Fermi
surface can be different from those described in previous
paragraph (see Fig.~1). But it is not difficult to see that the
conclusion of null charge Hall current remains valid.

It is possible to reveal the non-zero Berry curvature by adding a
very small (so that orbital quantization can be neglected)
perpendicular magnetic field to remove the point degeneracy. When
the chemical potential is within the Zeeman gap, only the lower
band is filled but now the point of high curvature is included in
the ${\vec k}$-integration in Eq.~(\ref{kubo}). Therefore, when
$\mu$ is within the Zeeman gap, the Hall conductance is roughly
equal to $e^2/(2h)$.\cite{culcer03}

In summary, crucial energy scales in the Rashba-Dresselhaus system
under an in-plane magnetic field are: the bottoms of the
spin-split bands $E_{\pm,{\rm min}}$, the degenerate energy $E_*$,
and the chemical potential $\mu$. All of these energies depend on
the applied magnetic field. By shifting the chemical potential
$\mu$ through $E_{+,{\rm min}}$ and $E_*$, the SHC could first be
enhanced (or reduced) and show a cusp, then become a
material-independent constant. Whether the SHC is enhanced or
reduced depends on, and can be controlled by, the direction of the
magnetic field. Finally, we caution that the charge Hall
conductivity depends subtly on the choice of eigenstates. As far
as we know, the SHC is not influenced by such subtle choice of
eigen-basis.

\acknowledgments The author is grateful to Dr. M.F. Yang for
numerous helpful discussions. He also thanks Dr. C.Y. Mou for
showing him S.K. Yip's paper in Ref.~\onlinecite{sheng96}. This
work is supported by the National Science Council under Contract
No. NSC 92-2112-M-003-011.

\appendix


\begin{thebibliography}{99}
\bibitem{below}
In Fig.~1(a), there are two contours since the bottoms of both
energy bands are lower than the degenerate energy and both bands
are populated. In some range of parameters, the bottom of the
upper band could coincide with the degenerate point. In that case,
there is only one contour in Fig.~1(a).
\bibitem{exclude}
The only exception is when $\alpha=\gamma$, which yields zero SHC.
\bibitem{gauge}
The eigenstates in Eq.~(\ref{states}) are related to those in
Ref.~\onlinecite{shen04} by a singular gauge transformation. The
Chern number would be different under such a singular gauge
transformation.
\end{thebibliography}
\end{document}